%% file: main.tex
\documentclass[conference]{IEEEtran}
\IEEEoverridecommandlockouts

\IEEEaftertitletext{\vspace{-1.2\baselineskip}}
\usepackage{cite}
\usepackage{amsmath,amssymb,amsfonts}
\usepackage{algorithmic}
\usepackage{graphicx}
\usepackage{textcomp}
\usepackage{xcolor}
\usepackage[
colorlinks=true,
urlcolor=blue,
citecolor=black,
linkcolor=black
]{hyperref}
\def\BibTeX{{\rm B\kern-.05em{\sc i\kern-.025em b}\kern-.08em
    T\kern-.1667em\lower.7ex\hbox{E}\kern-.125emX}}
\begin{document}

\title{Print-Aware Synthesis and Physical Design Methodologies for 3D-Printed Microfluidic Biochips
}


\author{\IEEEauthorblockN{Yushen Zhang}
\IEEEauthorblockA{
\textit{Technical University of Munich}\\
yushen.zhang@tum.de}
\and
\IEEEauthorblockN{Tsun-Ming Tseng}
\IEEEauthorblockA{
\textit{Technical University of Munich}\\
tsun-ming.tseng@tum.de}
\and
\IEEEauthorblockN{Ulf Schlichtmann}
\IEEEauthorblockA{
\textit{Technical University of Munich}\\
ulf.schlichtmann@tum.de}

}

\maketitle

\begin{abstract}
Microfluidic devices are widely used in diagnostics, chemical synthesis, and biological analysis, but their development often depends on complex fabrication and design processes. Resin-based three-dimensional (3D) printing has emerged as a promising alternative to conventional microfabrication because it enables low-cost, rapid prototyping of complex multi-layer structures. However, the practical realization of 3D-printed microfluidic biochips remains challenging due to manual and expertise-intensive design workflows, the rigid nature of commonly used printing materials, and fabrication inaccuracies such as over-curing that distort internal features and may block narrow channels. In this paper, we present a cohesive design automation framework for 3D-printed microfluidics that addresses these challenges across both device design and fabrication. The framework combines interactive design tools, automated synthesis methods for functional 3D microfluidic devices, techniques for developing low-cost 3D-printed mixers, and design-for-manufacturing strategies to improve print fidelity on low-cost resin printers.
\end{abstract}

\section{Introduction}
\vspace{-0.1cm}
Microfluidic biochips have emerged as a powerful technology in both research and industry, enabling the precise handling of small fluid volumes for applications including point-of-care diagnostics, drug delivery, chemical synthesis, and biological analysis \cite{han2009measuring, bilitewski2003biochemical, holmes2010application}. Traditionally, microfluidic fabrication has relied on processes such as soft lithography and polydimethylsiloxane (PDMS)–glass bonding. Although these methods can produce high-quality devices, they are typically labor-intensive, require specialized facilities, and depend heavily on cleanroom infrastructure \cite{love2001microscope}. These constraints limit accessibility and slow down iterative development.

\begin{figure}[b]
    \centering
    \vspace{-0.6cm}
    \includegraphics[width=\linewidth]{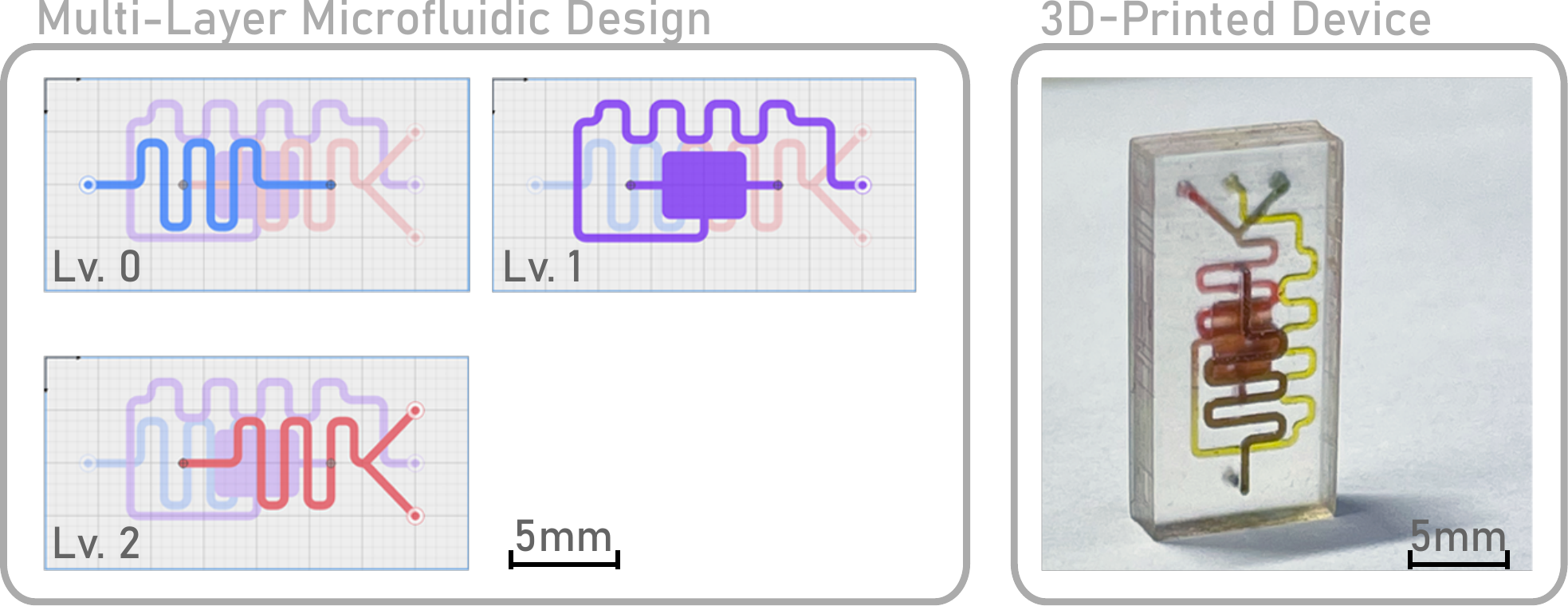}
    \vspace{-0.6cm}
    \caption{Multi-layer microfluidic chip fabricated by 3D printing \cite{lipcon}.}
    \label{fig:3dp-mf}
\end{figure}

In recent years, additive manufacturing, or three-dimensional (3D) printing, particularly resin-based methods such as stereolithography (SLA) and digital light processing (DLP), has attracted increasing attention as an alternative fabrication route for microfluidics. Compared with conventional approaches, 3D printing offers several compelling advantages: rapid prototyping, reduced infrastructure requirements, lower production costs, and the ability to fabricate complex three-dimensional and multi-layered structures that are difficult or even impossible to realize using planar manufacturing techniques. An example of such a multi-layer 3D-printed microfluidic chip is shown in Figure \ref{fig:3dp-mf}. The availability of low-cost SLA printers, many of which cost only a few hundred dollars, further strengthens the appeal of this approach, especially for laboratories and facilities with limited resources. In this sense, 3D printing has the potential not only to simplify fabrication, but also to democratize access to microfluidic technology. However, despite these advances in manufacturing, the design of 3D-printed microfluidic systems remains a major challenge. Developing such devices requires expertise spanning fluid dynamics, mechanical design, and fabrication-aware engineering. In practice, the design workflow still relies heavily on manual computer-aided design (CAD), which is time-consuming, difficult to scale, and increasingly error-prone as device complexity grows. This challenge becomes even more pronounced for multi-layer architectures, where the spatial arrangement of channels, chambers, and interconnections must satisfy both functional and manufacturing constraints. While earlier design automation tools \cite{10.1145/3768629, tseng2017columba, tseng2019cloud, sanka20193dmuf, 10.1145/3306492, amarasinghe2008nada, 10.1145/2660773} have improved the design of conventional planar microfluidic chips, they do not adequately address the distinctive requirements of 3D-printed systems.

A further limitation arises from the material properties of commonly available 3D-printing resins. Off-the-shelf materials for resin printing generally produce rigid and inelastic structures, in contrast to PDMS-based devices, which can exploit the elasticity of the substrate to implement active elements such as pneumatic valves and pumps. As a result, 3D-printed microfluidic devices often cannot rely on deformable components for active flow control and must instead achieve functionality primarily through the careful design of internal geometries and fluidic pathways. This strong dependence on geometry places greater demands on the design process: channel lengths, volumes, resistances, and timing relationships must be coordinated precisely to ensure correct fluid handling, synchronized reactions, and accurate dispensing behavior \cite{song2003microfluidic}.

At the same time, the physical realization of these designs is constrained by the fabrication inaccuracies inherent to resin-based printing. During printing, light penetration and scattering can cause polymerization outside the intended exposure region, a phenomenon commonly referred to as over-curing. This effect introduces geometric deviations from the nominal design and is especially problematic in narrow channels and in fine, complex internal features. In severe cases, over-curing can partially or completely block channels, rendering a device unusable.
\begin{figure}[!t]
    \centering
    \includegraphics[width=\linewidth]{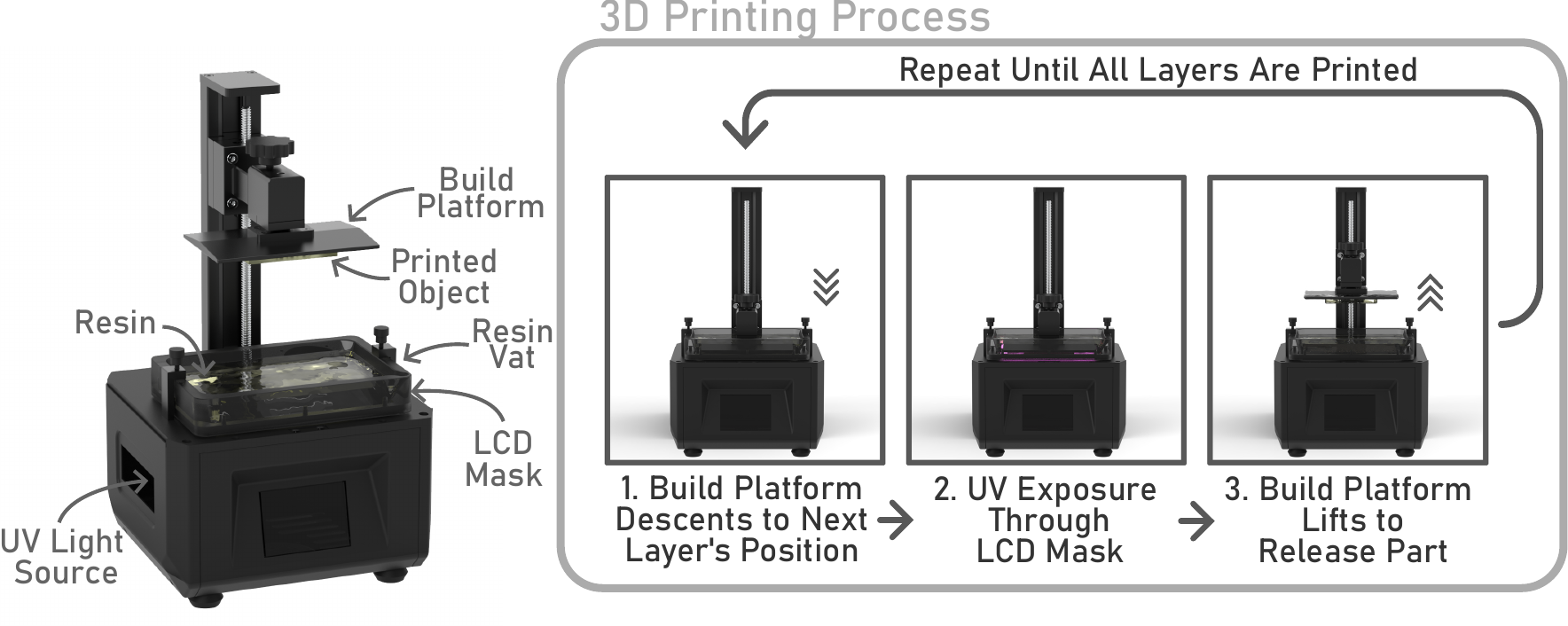}
    \vspace{-0.6cm}
    \caption{Illustration demonstrating how a resin-based SLA 3D printer creates an object.}
    \label{fig:printing-process}
    \vspace{-0.7cm}
\end{figure}
Taken together, these issues reveal a critical gap between the promise of 3D printing for microfluidics and its practical adoption. Although 3D printing has made fabrication more accessible, persistent challenges in design complexity, functional synthesis, material limitations, and fabrication fidelity continue to slow the broader dissemination of 3D-printed microfluidics. These issues are particularly significant in the context of low-cost fabrication. Hobby-grade printers are highly attractive for making microfluidics more accessible, yet they are also more susceptible to inaccuracies such as light-induced feature distortion. High-end printers may alleviate some of these problems, but they do so at the expense of affordability, thereby weakening one of the most compelling motivations for using 3D printing in the first place.

To address these challenges, this paper presents a cohesive design automation framework for 3D-printed microfluidics that unifies a series of interconnected methods across design and fabrication. Specifically, the framework spans interactive tools to lower the barrier to manual design of 3D microfluidic geometries, methods for automated synthesis of 3D-printable microfluidic devices, approaches for synthesizing low-cost 3D-printable mixing systems, and design-for-manufacturing techniques to improve fabrication fidelity on low-cost resin printers. By integrating interactive design support, constraint-driven functional synthesis, and machine-learning-assisted manufacturing compensation, the presented methodologies lower the barrier to entry for non-expert users and reduce the time and cost required to develop complex microfluidic systems. In doing so, it advances the goal of making 3D-printed microfluidics not only technically viable, but also broadly accessible.

\section{Background}

\subsection{3D Printing Technologies for Microfluidics}
The fabrication of microfluidic devices has historically been dominated by soft lithography and PDMS molding, which require specialized cleanroom environments and extensive manual labor. In response to these limitations, researchers have increasingly turned to diverse 3D printing techniques for microfluidic fabrication, including stereolithography (SLA), digital light processing (DLP), fused deposition modeling (FDM), and two-photon polymerization (2PP) \cite{waheed20163d}. Among these methods, resin-based printing approaches, particularly SLA and DLP, have recently emerged as highly accessible alternatives that offer high-resolution features at low cost. In these processes, a liquid photosensitive resin is selectively exposed to a UV light source, curing and solidifying the material layer-by-layer to build the desired 3D geometry, as illustrated in Figure \ref{fig:printing-process}. The 3D printing process can be viewed as dragging the object from the resin vat. 

\begin{figure}[t]
    \centering
    \includegraphics[width=\linewidth]{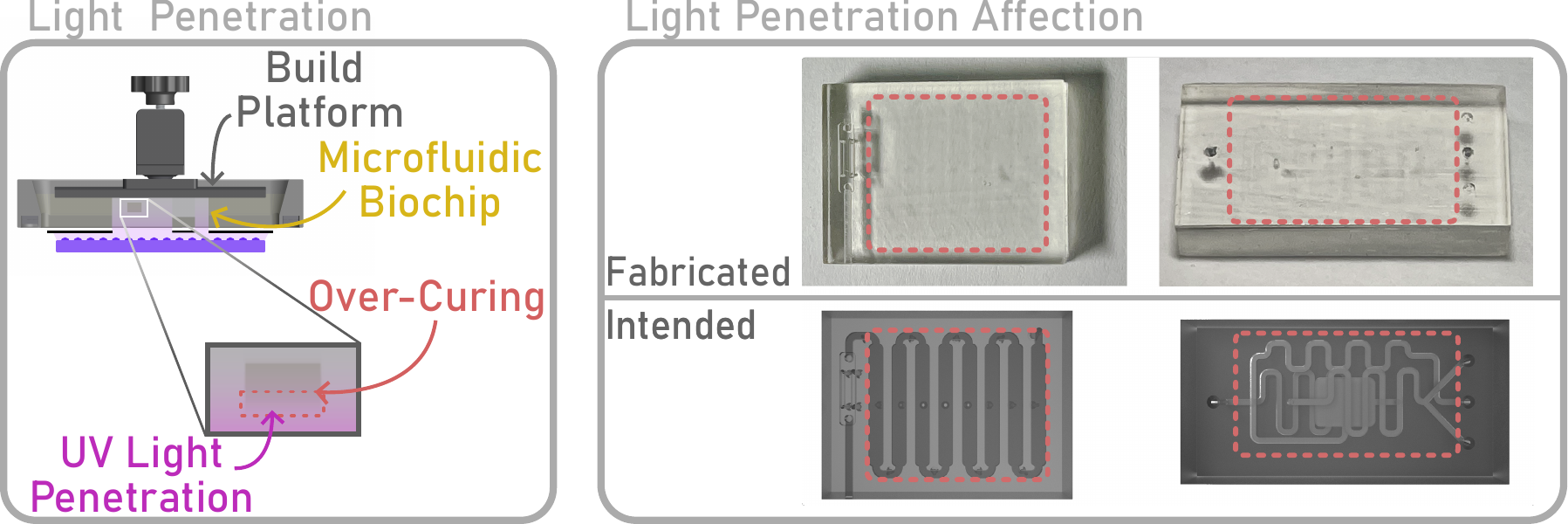}
    \caption{UV light penetration cures unintended areas during the printing process: (left) illustration of how UV penetrates through the printed layers; (right) light penetration affects the yield of 3D-printed microfluidic biochips.}
    \label{fig:lightpen}
    \vspace{-0.6cm}
\end{figure}
\subsection{Light Penetration and Fabrication Artifacts}
While SLA and DLP printing offer high resolution, the physical nature of photopolymerization introduces significant fabrication challenges, particularly for internal microfluidic voids. During the layer-by-layer printing process, UV light inevitably penetrates through the currently exposed layer and scatters into previously solidified layers or adjacent void regions, as illustrated in Figure \ref{fig:lightpen}(left). This phenomenon, known as light penetration or over-curing, causes unintended polymerization of the residual liquid resin trapped inside the microchannels.

\begin{figure*}[!t]
    \centering
    \includegraphics[width=\linewidth]{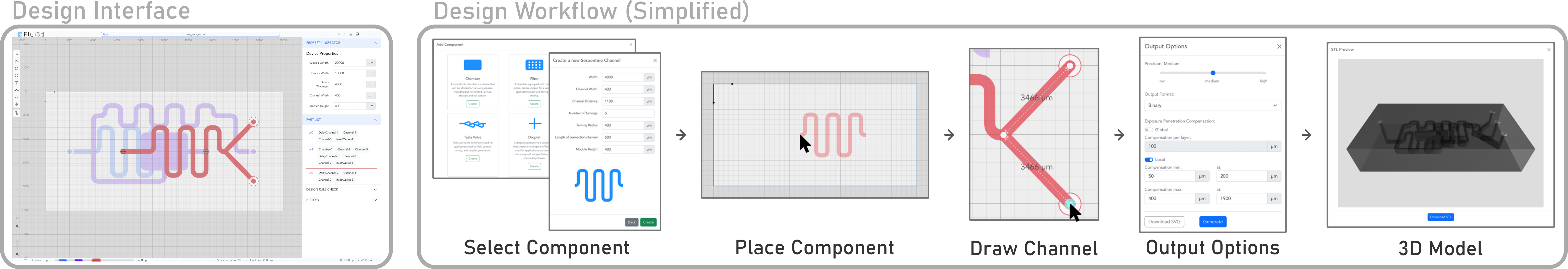}
    \vspace{-0.7cm}
    \caption{Flui3d simplifies the design of 3D-printable microfluidic biochips: (left) user interface of the design platform; (right) simplified design workflow illustrating the key design steps.}
    \label{fig:flui3d}
    \vspace{-0.7cm}
\end{figure*}

The extent of this light penetration depth is governed by the Beer-Lambert law, which dictates an exponential decay of light intensity as it travels through the resin. Neighboring structural features also reflect, refract, and scatter this light, creating localized variations in the over-curing effect. This cumulative UV exposure severely distorts the fabricated dimensions of the features, altering their intended hydraulic resistance and, in extreme cases, completely clogging the fluidic pathways, as shown in Figure \ref{fig:lightpen} (right). 

\subsection{Microfluidic Design Automation}

As the complexity of microfluidic networks increases, manual design using standard mechanical CAD software becomes a significant bottleneck. Designing a functional chip requires routing channels, calculating precise channel lengths to satisfy hydraulic resistance requirements, and ensuring that no components overlap. To address these challenges, Electronic Design Automation (EDA) tools for microfluidics have advanced rapidly in recent years, automating many of these labor-intensive tasks. Several state-of-the-art design automation tools, such as 3D{$\mu$}F \cite{sanka20193dmuf}, Cloud Columba \cite{tseng2017columba, tseng2019cloud}, Micado \cite{amarasinghe2008nada}, and Fluigi \cite{10.1145/2660773}, have significantly facilitated the design of continuous-flow microfluidic devices. For instance, 3D{$\mu$}F provides an interactive environment for planar chip design, while Cloud Columba supports synthesis for continuous-flow mLSI systems. These tools can automatically generate 2D layouts, vector graphics, or CAD scripts compatible with conventional fabrication methods. However, despite their capabilities, most existing microfluidic design automation solutions are fundamentally limited to 2D architectures, often relying on layered flow and control designs. As a result, they are not directly applicable to the design of microfluidic devices intended for 3D printing. This limitation highlights the growing need for domain-specific CAD and design automation tools tailored to fully leverage the advantages of 3D-printed microfluidics.
This lack of integrated, 3D-capable design synthesis tools forces researchers into tedious, iterative manual design cycles to satisfy specific timing, volumetric, or mixing constraints.

\section{Interactive Design Automation for Multi-Layer Architectures}
As the field of microfluidics grows from planar, single-layer devices to complex, multi-layer 3D architectures, the limitations of traditional mechanical CAD software become increasingly apparent. Standard CAD tools are not optimized for fluidic routing, requiring manual, tedious adjustments for every channel bend, layer transition, and component placement. To overcome these infrastructural bottlenecks, we developed a foundational web-based interactive design automation platform---\textit{Flui3d}---specifically tailored for 3D-printed microfluidics \cite{zhangFlui3d}. Without installing any software, Flui3d can be easily accessed via the web at \url{https://flui3d.org}. 
\begin{figure*}[t]
    \centering
    \includegraphics[width=\linewidth]{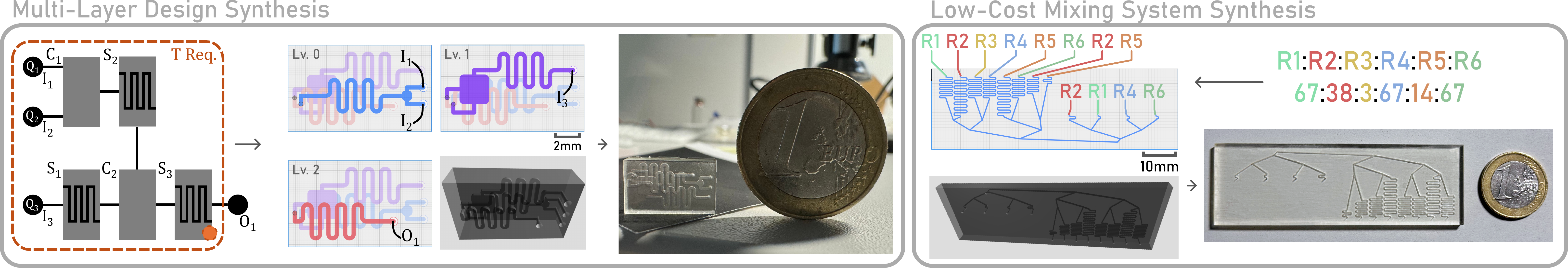}
    \vspace{-0.6cm}
    \caption{Automated design synthesis examples: (left) multi-layer architecture generation under timing constraint, from abstract schematic to fabricated chip; (right) passive 3D-printed mixer synthesis with user-defined multi-reagent ratios.}
    \label{fig:synth}
    \vspace{-0.6cm}
\end{figure*}

\textit{Design user interface.} Flui3d provides a web-based, WYSIWYG (What-You-See-Is-What-You-Get) design environment in which users can create microfluidic layouts through a familiar 2D interaction paradigm, while the tool automatically generates the corresponding 3D geometry and fabrication-ready output. Figure \ref{fig:flui3d} shows the user interface and the key design steps. The platform includes a parameterized component library, support for multi-layer device design, and export options including STL, SVG, and a human-readable JSON representation for sharing and reuse.

\textit{Parameterized component library.} A central element of Flui3d is its parametric component library, which abstracts microfluidic design away from low-level geometry construction toward domain-specific building blocks. Instead of manually defining channels and volumes in 3D space, users compose devices from functional primitives such as channels, chambers, mixers, Tesla valves, ports, and inter-layer connections. Each component is parameterized, allowing users to adjust dimensions and properties while maintaining internal consistency and connectivity constraints. This approach encapsulates domain knowledge directly into the design representation, enabling rapid iteration and reducing the likelihood of invalid geometries. As a result, the platform shifts the design task from geometric modeling to the configuration and composition of microfluidic functionality.

\textit{Simple design workflow.} Complementing the component-based abstraction, Flui3d structures the design process as a guided workflow that reflects the logical steps of microfluidic device development. Users first define global device dimensions and default geometric parameters, then create and organize layers, place and parameterize components, add ports, define inter-layer connections (vias), and route channels between elements. The workflow is completed by design rule checking and automated generation of printable output. In this way, Flui3d turns the design of multi-layer 3D-printed microfluidics into a structured sequence of high-level operations rather than a manual 3D modeling task. For example, the three-layer mixing device from Figure \ref{fig:3dp-mf} could be taken from initial setup to a print-ready design in less than five minutes, demonstrating the efficiency of this interaction model.

\section{Design Synthesis for 3D-Printed Microfluidics}

\subsection{Multi-Layer Architectural Synthesis}

A central bottleneck in 3D-printed microfluidics is the lack of synthesis tools that can jointly handle multi-layer physical layout generation and user-defined fluid-delivery behavior. In passive chips, this challenge is particularly acute because functional correctness is encoded directly in geometry. Designers must simultaneously satisfy spatial constraints---such as placement, routing, non-overlap, and layer usage---and hydraulic constraints, including flow, timing, and filling volume. While manual CAD workflows can meet these requirements for simple devices, they become highly iterative and error-prone as designs scale to multiple layers and involve tightly coupled timing and volume targets. To address this problem, we introduced \textit{3M-DeSyn} \cite{zhang20253m}.

\textit{Input model and specification.} 3M-DeSyn begins from a high-level schematic that specifies functional components (e.g., mixers, chambers, sinks), connectivity, and available inlets. The user can add optional directives, including preferred layer counts, placement hints for critical modules, target inlet flow conditions, synchronization constraints (equal arrival times), shifted-arrival constraints (prescribed delays between paths), and volumetric constraints (absolute target volumes or target volume ratios at sinks). This representation decouples functional intent from final geometry and enables automated exploration of physical realizations.

\textit{Graph abstraction and physical embedding.} The schematic is mapped to a graph whose vertices represent components and whose edges represent channels to be synthesized. Physical embedding introduces geometric variables for node locations, layer assignments, channel lengths, and routing segments. Additional constraints enforce manufacturable spacing, prevent overlap among components and channels, and guarantee legal inter-layer transitions. This formalization allows layout synthesis to be solved together with hydraulic design rather than as separate sequential steps.

\textit{Hydraulic and temporal formulation.} Under laminar-flow assumptions, 3M-DeSyn couples geometry to behavior using resistance- and flow-based relationships, together with a volume-to-time relation ($T=V/Q$). For each constrained path, timing requirements are encoded as equalities or inequalities with tolerances. Simultaneous-arrival constraints are modeled by equalizing effective path delays, while shifted-arrival constraints are implemented by introducing calibrated volume offsets that produce prescribed delay differences under the assigned flow conditions. In this way, routing decisions directly influence whether temporal objectives are satisfied.

\textit{Volumetric control.} For metering and ratio synthesis, the framework converts desired output volumes into path-level constraints that connect channel geometry, flow distribution, and filling times. When users specify relative targets (e.g., ratio-based dispensing), 3M-DeSyn enforces proportional relationships among effective delivered volumes. This is particularly important for passive reagent preparation, where concentration control depends on maintaining robust volume ratios without active valves or pumps.

\textit{Optimization objective and solver strategy.} The full design is cast as a constrained optimization problem. The objective combines compact footprint, feasible routability, and low routing complexity, while satisfying all hard geometric and functional constraints. In practice, this yields a coordinated placement-and-routing synthesis process in which channel dimensions are not post-hoc tuned manually, but are synthesized as part of the optimization itself. The resulting solution is then converted automatically into 3D geometry for fabrication.

\textit{Experimental outcomes.} 3M-DeSyn was evaluated on five test cases spanning different combinations of timing, volumetric, and multi-layer constraints. Compared with a manual baseline workflow (schematic analysis, iterative placement/routing, and hand-tuned channel dimensions), 3M-DeSyn reduced design time from hours to seconds/minutes: baseline estimates ranged from 5\,h\,07\,m to 20\,h\,54\,m, whereas 3M-DeSyn completed synthesis in 2\,s to 3\,m\,40\,s. The synthesized chips were additionally fabricated on a low-cost hobby SLA setup (Anycubic Photon D2 with plant-based clear resin), confirming that the automatically generated geometries are not only computationally valid but also practically printable. An example is shown in Figure \ref{fig:synth}(left).

\subsection{Low-Cost Mixing System Design Synthesis}

Complementing 3M-DeSyn, \textit{\(\mu\)-MM} \cite{11420367} focuses on a different but closely related challenge: synthesizing passive 3D-printed mixers that deliver \emph{user-defined multi-reagent ratios} under a \emph{single constant pressure source}. The central motivation is to avoid external complexity (multiple pumps, controllers, or valves) while still achieving ratio-specific mixing that is accurate enough for assay and diagnostic workflows. In contrast to many passive mixer designs that primarily optimize homogeneity for symmetric or fixed settings, \(\mu\)-MM targets arbitrary ratio specifications and explicitly incorporates manufacturability for low-cost resin printing.

\textit{Problem formulation.} Given a target ratio vector
\(T = r_1 : r_2 : \cdots : r_n\), where each \(r_i\) denotes the required volumetric contribution of reagent \(i\), the goal is to automatically generate a complete mixer layout (inlets, junction topology, channel dimensions, and 3D geometry) such that the outlet mixture follows the requested ratio under one shared pressure input. This framing intentionally shifts complexity from hardware control to algorithmic design synthesis.

\textit{Ratio decomposition and mix-flow tree construction.} The first stage deterministically decomposes the target ratio into additive sub-ratios and maps them into a hierarchical \emph{mix-flow tree}. In this tree, leaves represent inlet branches, internal nodes represent mixing junctions, and the root corresponds to the final outlet. Decomposition increases flexibility in flow allocation: each reagent contribution can be represented by multiple sub-streams that are later interleaved spatially. This strategy allows the method to preserve exact aggregate ratio contributions while exposing additional degrees of freedom for improving the downstream mixing pattern.

\textit{Spectrum-aware arrangement for passive mixing quality.} Beyond ratio correctness, \(\mu\)-MM introduces a spectrum-oriented arrangement of decomposed sub-flows. Intuitively, by distributing and reordering subcomponents in the tree, the outlet can exhibit more distinct laminar sub-streams (referred to as wider spectrum width), which supports better passive homogenization after merging. This is an important design choice because passive systems cannot rely on active perturbation; therefore, flow ordering and geometric interleaving become key levers for practical mixing performance.

\textit{Layout generation and hydraulic modeling.} In the second stage, the abstract tree is translated into a compact physical layout using block-based placement and routing constraints. The method further transforms tree levels when beneficial, so that selected lower-level nodes are elevated to reduce excessive path growth and improve area efficiency. Channel dimensions are then determined via the electronic-hydraulic analogy, where each edge receives a hydraulic resistance realized through channel length assignment. Under a common pressure boundary, these resistive paths enforce the intended relative flow rates so that outlet composition tracks the original target ratio.

\textit{Constraint optimization and geometry export.} The physical realization is solved as a constrained optimization problem that balances routability and compactness while preserving the synthesized flow structure. The output is not merely a logical network; it is a fabrication-ready geometric design that can be exported in standard 3D format and sent directly to slicers/printers.

\textit{Experimental validation.} The approach was evaluated on eight case studies (four real-world-inspired and four synthetic ratio targets). Generated layouts were fabricated using a hobby-grade resin printer (Elegoo Mars 5). An example is shown in Figure \ref{fig:synth}(right). To validate ratio accuracy, a  COMSOL-based laminar-flow/solute-transport simulation was performed, assigning distinct inlet concentrations to reagents so outlet composition errors could be quantified against theoretical targets.
Across the four real-world scenarios, simulated outlet concentrations deviated by less than 0.01\,mol/L from target values. Over all eight cases, reported deviations remained small (from \(-0.016\) to \(+0.009\)\,mol/L), while synthesis runtime stayed in the millisecond range (108--145\,ms). Chip footprints and total channel lengths remained moderate relative to a custom baseline, highlighting that \(\mu\)-MM achieves an effective balance: slightly narrower mixing spectrum in some cases, but substantially improved compactness and manufacturability with lower channel complexity and hydraulic burden.

\section{Design-for-Manufacturing for 3D-Printed Microfluidics}

\subsection{Manual Light-Penetration-Compensation}

A core manufacturability bottleneck in resin-based 3D-printed microfluidics is the unintended curing caused by light penetration through previously printed material. This artifact is especially severe for enclosed small channels and multi-layer layouts, where over-exposure can narrow or fully block critical flow paths. We introduced practical Design-for-Manufacturing (DFM) compensation strategies that is integrated directly into the design-output stage of Flui3d, enabling reliable fabrication on hobby-grade SLA/LCD platforms without requiring custom resins or printer hardware modifications.

\textit{Problem focus and DFM objective.} While layout synthesis can generate functionally correct geometries, printability is not guaranteed when optical side effects distort fine features. The DFM module therefore treats fabrication fidelity as a first-class design objective: before STL export, the geometry is transformed to counter expected light-induced over-curing. The goal is to preserve intended void regions and keep microchannels open after printing, especially in compact and vertically integrated devices.

\begin{figure}[t]
    \centering
    \includegraphics[width=\linewidth]{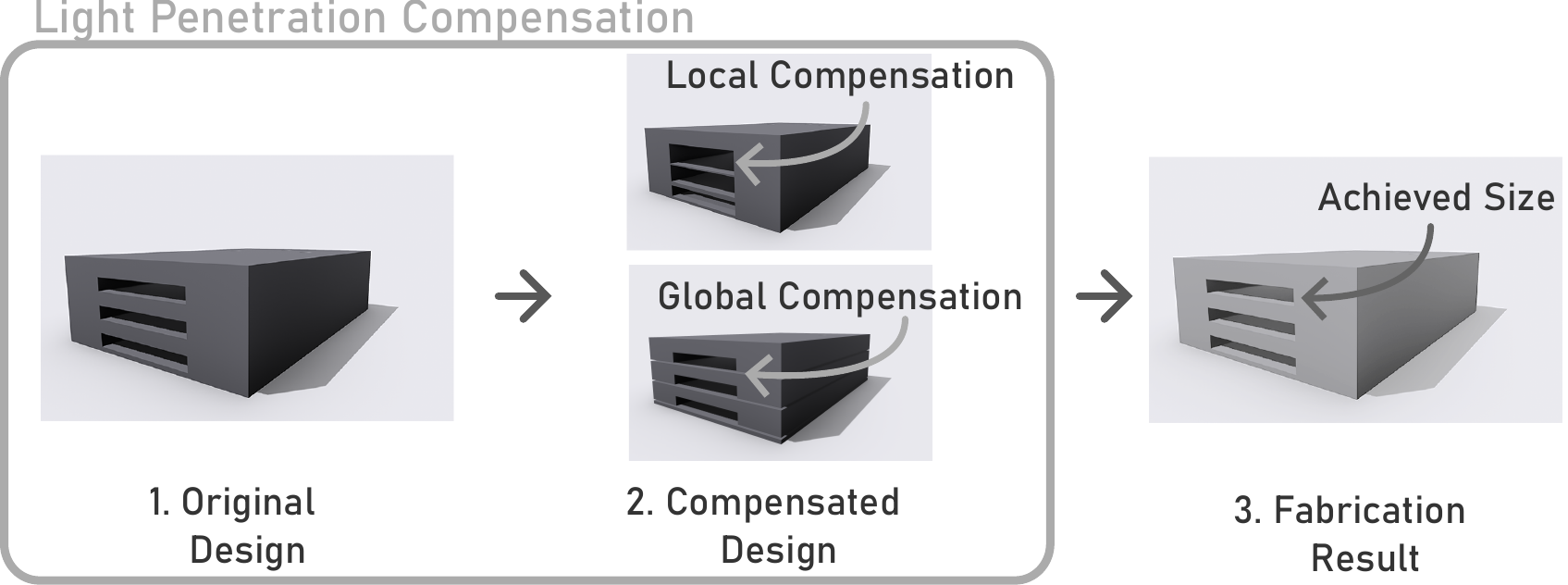}
    \vspace{-0.6cm}
    \caption{The principle of our proposed light penetration compensation.}
    \label{fig:lpc}
    \vspace{-0.6cm}
\end{figure}

\textit{Two complementary compensation strategies.} The framework introduces two compensation strategies that can be used independently or jointly. Figure \ref{fig:lpc} illustrates the principle.
First, \emph{local compensation} adds feature-specific height offsets that vary with layer depth, reflecting the fact that received light dose differs by position within the print.
Second, \emph{global compensation} inserts a user-defined blank exposure height after feature layers, increasing optical separation and reducing unintended polymerization in nearby void regions.
This local-plus-global decomposition provides a flexible mechanism: local compensation addresses depth-dependent exposure nonuniformity, while global compensation provides additional robustness for difficult print settings.

\textit{Physics-guided local compensation.} Local compensation is derived from a Beer--Lambert-style attenuation assumption in which light intensity decays exponentially with distance from the source. From this model, cumulative exposure at a given height is mapped to a compensation amount \(C(z)\), yielding a dynamic layer-wise offset profile rather than a single uniform margin. In practice, this allows deeper or more vulnerable regions to receive stronger compensation automatically, which is particularly useful for multi-layer channel stacks.

\textit{User-parameterized calibration workflow.} To avoid requiring full optical characterization of each printer-resin combination, the method uses a practical calibration interface based on four user inputs: minimum and maximum desired compensation values (\(C_{\min}, C_{\max}\)) and their corresponding height locations (\(Z_{\min}, Z_{\max}\)). These anchor points are used to approximate model coefficients and generate the full compensation curve applied across layers. A reference calibration design is provided so users can print, inspect, and iteratively tune these settings for their own machine, resin, and exposure conditions.

\textit{Role of global compensation in hard regimes.} Global compensation is particularly valuable when local compensation alone is insufficient, such as low-power printers that require longer exposure times per layer or designs with dense multi-layer interactions. By adding blank height between feature layers, the method increases the optical path and reduces cross-layer curing risk. The work also notes a practical tradeoff: excessive global offset can reduce structural stability, so this parameter must be tuned alongside local settings.

\vspace{-0.2cm}
\subsection{ML-Based Light-Penetration-Compensation}
\vspace{-0.1cm}
We further advanced the design-for-manufacturing direction with \textit{LiPCon} \cite{lipcon} by introducing a \emph{physics-aware, spatially resolved} light-penetration compensation method. In contrast to prior uniform-compensation approaches, LiPCon explicitly models the fact that over-curing depends not only on feature depth but also on nearby geometry and local optical context. The central goal is therefore to predict \emph{location-specific} compensation values that preserve intended feature dimensions in complex multi-layer designs.

\textit{Dataset contribution and data pipeline.} This work contributes a dedicated dataset quantifying discrepancies between design and fabrication in resin-based multi-layer microfluidics. The dataset was constructed from calibration structures with controlled variations in feature geometry and spatial context, followed by measurement of fabricated dimensions under fixed print settings. This produced 2,488 data points, providing supervised targets for compensation learning.

\textit{Physics-aware preprocessing.} To improve learnability and physical consistency, LiPCon introduces preprocessing that embeds light-propagation principles into the model input representation. Specifically, geometric features are weighted based on cumulative light passage and attenuation effects (including absorption/scattering-related decay), so the model receives physically meaningful context instead of only raw geometric descriptors. This hybrid strategy combines domain priors with machine learning, reducing reliance on purely black-box fitting.

\textit{Spatially resolved machine-learning model.} LiPCon then trains a neural predictor to output local compensation values for different regions/features in a design. Unlike uniform compensation, this allows stronger correction where local over-curing risk is high and milder correction where risk is lower. In practice, the learned compensation map is injected into the DFM stage before fabrication, producing a compensated 3D model whose printed dimensions are closer to the original CAD targets.

\textit{Training setup and predictive performance.} The dataset was split into training/validation/test subsets (1,727/315/434). Training used Adam with MSE loss, dropout-based regularization, and early stopping. The model converged in fewer than 40 epochs, achieved about 90.01\% training accuracy, and showed tight test-error behavior: all reported \(L_1\) errors were below 0.45 with an average around 0.11, indicating stable generalization for local compensation prediction.

\textit{Fabrication validation and benchmarking.} For end-to-end validation, multiple random multi-layer designs were fabricated and compared against the prior state-of-the-art baseline method. Across representative microfluidic features (e.g., narrow channels, serpentine regions, and cavity-like structures), LiPCon produced fabricated dimensions consistently closer to design targets. For six cases printed on the same printer-resin setup as training data, dimensional errors were reduced by at least 87.64\% relative to baseline, with final deviations only 1.25\%--5.5\%. The method was also applied to an additional printer-resin pair without retraining, where deviations remained within 5.75\%--17.00\% and still outperformed the baseline, demonstrating practical robustness and transfer potential.

\section{Conclusion}
This paper presented a unified automation framework for 3D-printed microfluidics that addresses the full path from design to fabrication. By combining interactive design support, constraint-driven synthesis for passive fluidic functions, low-cost mixer synthesis, and compensation methods for light-penetration artifacts, the framework reduces manual effort while improving print reliability on resin 3D printers. Experimental results across representative designs show substantial reductions in design time and strong agreement between fabricated and target dimensions. Overall, this work helps bridge the gap between the accessibility of low-cost 3D printing and the practical requirements of reliable microfluidic device development.

\bibliographystyle{IEEEtran} 
\input{output.bbl}

\end{document}